\documentclass[referee]{raa}            

\usepackage{graphicx,times}             
\usepackage{natbib}
\usepackage{amssymb,amsmath}
\bibpunct{(}{)}{;}{a}{}{,}
\usepackage[]{hyperref}
\hypersetup{colorlinks = true, linkcolor = green, anchorcolor = red, citecolor = blue, filecolor = red, pagecolor = red, urlcolor = red}

\usepackage[bottom = 0.5in]{geometry}

\newcommand{\tgt}{J0935.3+0901}
\newcommand{\gr}{$\gamma$-ray}
\newcommand{\fermi}{\textit{Fermi}}
\newcommand{\xmm}{\textit{XMM-Newton}}

\begin{document}

   \title{X-ray and Radio Studies of the candidate Millisecond Pulsar Binary 4FGL J0935.3+0901
}

   \volnopage{Vol.0 (20xx) No.0, 000--000}      
   \setcounter{page}{1}          

   \author{Dong Zheng
      \inst{1}
   \and  Zhong-Xiang Wang
      \inst{1,2,3}
   \and  Yi Xing
      \inst{3}
   \and Jithesh Vadakkumthani
     \inst{4,5}
   }

   \institute{Department of Astronomy, School of Physics and Astronomy,Yunnan University, Kunming 650091, China
             \\
        \and
	Key Laboratory of Astroparticle Physics of Yunnan Province, 
	Yunnan University, Kunming 650091, China
	     {\it wangzx20@ynu.edu.cn}\\
        \and
       Shanghai Astronomical Observatory, Chinese Academy of Sciences, 80 Nandan
 Road, Shanghai 200030, China\\
 	\and 
	Inter-University Centre for Astronomy and Astrophysics, PB No.4, Ganeshkhind, Pune-411007, India\\
	\and Department of Physics, University of Calicut, Malappuram-673635, Kerala, India\\
\vs\no
   {\small Received~~20xx month day; accepted~~20xx~~month day}}

\abstract{ 4FGL J0935.5+0901, a $\gamma$-ray source 
recently identified as a candidate redback-type millisecond pulsar 
binary (MSP), shows an interesting feature of having
double-peaked emission lines in its optical spectrum.
The feature would further
suggest the source as a transitional MSP system in the sub-luminous disk state.
We have observed the source with \xmm\ and Five-hundred-meter Aperture 
Spherical radio Telescope (FAST) at X-ray and radio frequencies respectively
for further studies. From the X-ray observation, a bimodal count-rate 
distribution, which is a distinctive feature of the transitional MSP systems, 
is not detected, while the properties of X-ray variability and power-law 
spectrum are determined for the source. These results help establish 
the consistency of it being a redback in the radio pulsar state. However no
radio pulsation signals are found in the FAST observation, resulting an 
upper limit on the flux density of $\sim 4\,\mu$Jy. Implications of these
results are discussed.
}

   \authorrunning{Zheng et al.}            
   \titlerunning{Candidate MSP binary 4FGL J0935.3+0901}  

   \maketitle

%
%
\section{Introduction}           
\label{sect:intro}

Millisecond pulsars (MSPs) form from low-mass X-ray binaries (LMXBs), in which
a neutron star gains angular momentum by accreting mass that is transferred 
from a Roche-lobe filling companion star \citep{bv91}. As one type of
the end products of binary evolution, a class of so-called 
`redback' MSP binaries has recently been recognized \citep{rob13} and 
gained attention.
These close MSP binaries contain a $\ge 0.1 M_{\odot}$ companion star, 
constituting a class sample that helps reveal not only the LMXB evolution 
processes
\citep{che+13,bdh14} but also intriguing physical properties of compact star
binaries (\citealt{str+19} and references therein). Moreover, two of redbacks
(PSR~J1023+0038 and XSS~J12270$-$4859; \citealt{arc+09,bas+14})
have been identified as transitional MSP (tMSP) systems 
and several as candidates (e.g., \citealt{mil+20} and references therein):
these systems can switch between the states of having an accretion disk 
(a so-called sub-luminous disk state) and being a disk-free, rotation-powered 
radio binary pulsar. Interesting multi-wavelength properties have been seen in
the tMSPs; for details, see, e.g., \citet{pat+14,sta+14,tak+14,sha+15,mil+20}.
Generally when in the disk state, no radio pulsed emission is detectable;
while in the disk-free state, tMSPs appear like regular redbacks.

The all-sky survey capability of the Large Area Telescope (LAT) onboard
{\it the Fermi Gamma-ray Space Telescope (Fermi)} has greatly enabled the
identification of redbacks. Candidate MSPs including redbacks can be selected 
from unidentified sources detected by \fermi\ LAT, and followup multi-wavelength 
observations will allow identification of redbacks by comparing obtained 
properties to those of the known systems. 
Of course, detection of millisecond pulsations
and orbitally periodic variations is required in order to identify a redback
(otherwise a source would be considered as a candidate).
Recently we have identified a candidate redback using the approach, for which 
the source in the \fermi\ LAT fourth source catalog (4FGL; \citealt{4fgl20}),
4FGL\,\tgt, was targeted and a 2.5-h binary was found \citep{wan+20}. 
Considering the observational facts,  the source's $\gamma$-ray
variability and curved spectrum, a faint X-ray counterpart (to the binary), and 
double-peaked optical emission lines (H$_{\alpha}$, H$_{\beta}$, and He~I)
of the binary (likely indicating the presence of a 
 $\sim 10^{10}$\,cm size accretion disk),
we have suggested 4FGL\,\tgt\ as a candidate redback and also likely
a transitional system in the sub-luminous disk state.

The tMSPs  in the disk state have distinctive properties. One is their 
bimodal count-rate
distributions at X-rays \citep{pat+14,bh15,cot+19,mil+20}. 
Between two distinguishable count-rate levels, a tMSP would switch quickly.
Whether 4FGL\,\tgt\
has this property can be checked with a targeted X-ray observation.  The 
ratios of 0.5--10\,keV X-ray fluxes to 0.1--100\,GeV \gr\ ones for
the tMSPs and candidates are 
$\sim$0.3--0.4, significantly higher than those of redbacks 
(e.g., \citealt{mil+20}), but the previously estimated flux ratio for
4FGL\,\tgt\ (at 0.3--10\,keV X-ray band) was only $\sim$0.02. This inconsistency thus raises a question
about the nature of 4FGL~\tgt\ and also about how to explain its double-peaked
emission line feature. In addition to being the indicator of an accretion
disk, the other possibility that the emission feature might reflect 
the intrabinary interaction processes was discussed in \citet{wan+20}.
The latter possibility would suggest the source as a regular redback, and then
a pulsar's pulsed emission might be detectable (note that there are candidate
redbacks in which radio pulsars are not detected; see, e.g., \citealt{str+19}).
A deep radio observation may clarify the nature of the source.

In order to further study this source, understanding its properties and
identifying its nature, we applied for \xmm\ and Five-hundred-meter Aperture
Spherical radio Telescope (FAST) times for X-ray and radio observations of
the source respectively. In this paper, we report the results from the 
observations.


\begin{figure}
   \centering
   \includegraphics[width=0.8\textwidth, angle=0]{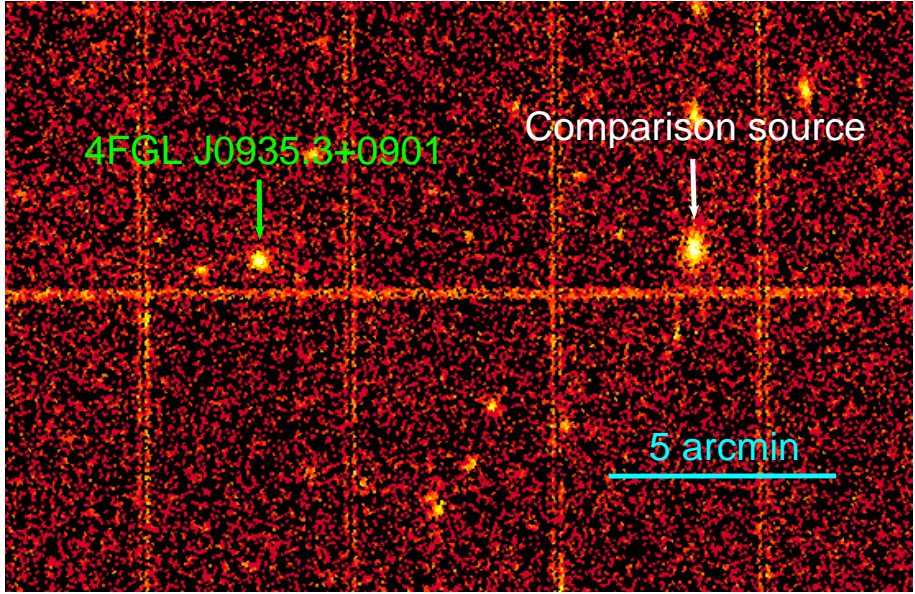}
   \caption{\xmm\ pn image of the source field, with \tgt\ and a comparison 
	source indicated.
	\label{fig:field}}
   \end{figure}

\section{Observations and Data Analysis}
\subsection{\textit{XMM-Newton} Observation}
\label{sec:XMM}

We observed 4FGL~\tgt\ on 2020 November 12 with the European Photon Imaging 
Camera (EPIC) onboard the \xmm\ space telescope. The total on-source observing
time was $\simeq$10.7\,ks. The observation (ObsID 0860350101) was
conducted in the full-frame mode with the thin optical filter used. 
The data were processed with the Science Analysis Software (SAS) data
reduction package (version~18.0.0). Full-field background light curves
in 10--12\,keV were extracted for the purpose of checking any presence of
particle flares in the pn and MOS1/MOS2 data. No background flares were seen
in our data.

\tgt\ was detected at the position of R.A.=$09^h35^m20^s.9$, 
Decl.=$09^{\circ}00^{'}38^{''}.27$, with a 1$\sigma$ uncertainty of 0$^{''}.50$
(the 1$\sigma$ systematic uncertainty for the position is 1$^{''}.5$). The
source was faint, with flux approximately equal to that estimated from 
the {\it Neil Gehrels Swift Observatory (Swift)} detection \citep{wan+20}. 
In order to study its X-ray 
variability, we chose a source detected in the field as a comparison source.
This source, at R.A.=$09^h34^m44^s.9$ and Decl.=$09^{\circ}03^{'}54^{''}.72$
(with a 1$\sigma$ uncertainty of 0$^{''}.40$), was only slightly brighter than 
our target. We note that this source could be a quasar based on its 
power-law spectrum (photon index $\Gamma\simeq 2.57$) and the comparison 
of its optical and X-ray fluxes ($V\simeq 18.8$ and unabsorbed flux 
$\simeq 2.8\times 10^{-13}$\,erg\,s$^{-1}$\,cm$^{-2}$, respectively; 
see, e.g., \citealt{he+19}). However we still used this source because
other sources in the field were too faint
and no significant variability was seen
in our following analysis of this comparison source.


To extract photons of \tgt\ and the comparison source,
a circular region with radius 24$^{''}$ was respectively used. 
The same-size background regions, local to the two sources respectively, 
were used. The best-quality data with FLAG=0 and PATTERN$\leq 4$ for pn
and PATTERN$\leq$12 for MOS1/MOS2, plus \#XMMEA\_EP and \#XMMEA\_EM
(for pn and MOS1/MOS2, respectively), were selected and used. For variability
analysis, we used the SAS tasks {\tt evselect} and {\tt epiclccorr} to extract
light curve data of the two sources. The final light curves were 
background-subtracted, produced by combining the data of MOS1, MOS2, and pn.
\begin{figure}
   \centering
   \includegraphics[width=0.8\textwidth, angle=0]{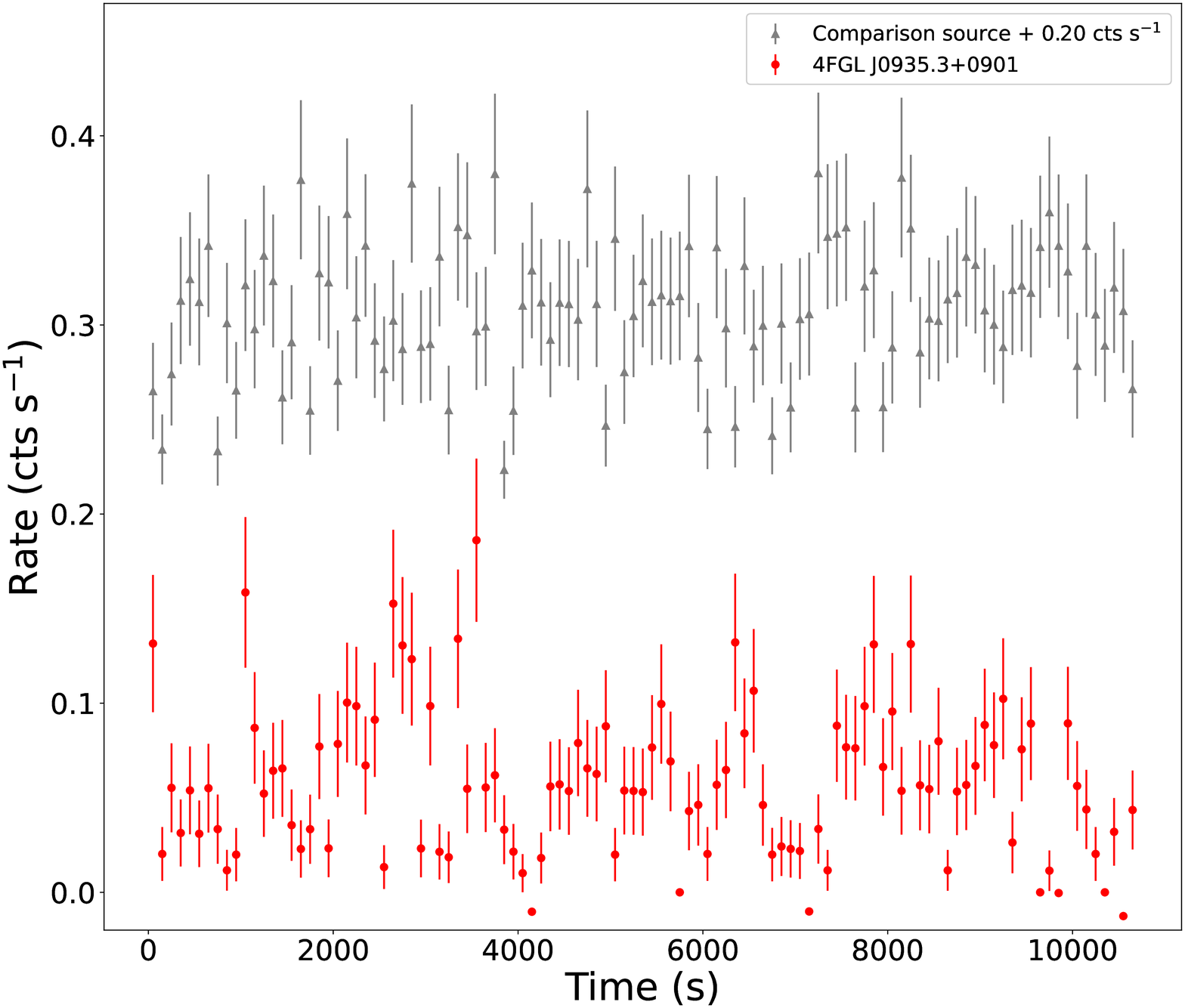} \\
   \caption{100-s binned, background-subtracted X-ray light curve of 4FGL~\tgt\ in 0.3--10\,keV.
	The same light curve of the comparison source is also shown. For
	clarity, the latter light curve is upshifted by 0.2\,cts\,s$^{-1}$.
	\label{fig:lc}}
   \end{figure}

We also obtained an EPIC spectrum of \tgt, combined from MOS1, MOS2, and pn
data using {\tt epicspeccombine}. Each spectral bin contains at least 20 counts.

%

\subsection{FAST Observation}

We observed 4FGL~\tgt\ with FAST \citep{nan06} on 2020 August 23, lasting
20\,min from 12:35:00 to 12:55:00 (UTC). The
Pulsar Backend system was used, which has 4096 channels in the waveband of
1.05--1.45\,GHz.  A sampling of 49.152\,$\mu$s was chosen in order to detect 
millisecond pulsation signals. The observation was conducted with
the telescope in the tracking mode.

The periodicity searching was performed with the pulsar processing software 
PRESTO \citep{ran01,rem+02,rce+03}. The dispersion measure (DM) from the 
Earth to 4FGL J0935.3$+$0901 is likely in a range of from 15.0 0 to 
28.31\,pc\,cm$^{-3}$, suggested by the estimated distance range for 
the source (0.76--2.4 kpc; see below Section~\ref{sec:disc}) based on 
the YMW16 electron-density 
model \citep{ymw17}. The time series were dedispersed over DMs in the likely
range using a DM step of 0.1\,pc\,cm$^{-3}$. 
We applied the acceleration search to retain sensitivity for possible binary 
pulsar by setting the zmax value to be 300, and folded the dedispersed time 
series on the candidate signals identified by the {\tt ACCEL\_sift} 
subroutine. No convincing pulsation signals were detected.

\section{Result}
\label{sect:Results}

\subsection{X-ray variability}

A distinctive feature of the tMSPs in the disk state is their bimodal 
distribution of X-ray count
rates with fast (as short as tens of seconds) switching timescales between 
the two count-rate levels (e.g., \citealt{pat+14}). While the faintness
of the source hampered our analysis to some degree, we did not see such a 
feature in \tgt. In Figure~\ref{fig:lc}, a light curve binned in 100\,s in
0.3--10\,keV is 
shown as an example. No phenomenon that the count rates stay at a high
level most time and can jump to a $\sim$10 times lower level in 10--100\,s
(see \citealt{pat+14,bh15,cot+19}) is seen.
We tested different time bins, such as 10\,s and 50\,s, but the results
were nearly the same.
Therefore neither two count-rate levels nor fast switching are evident in
X-ray emission of the source.
 
\begin{figure}
   \centering
   \includegraphics[width=0.89\textwidth, angle=0]{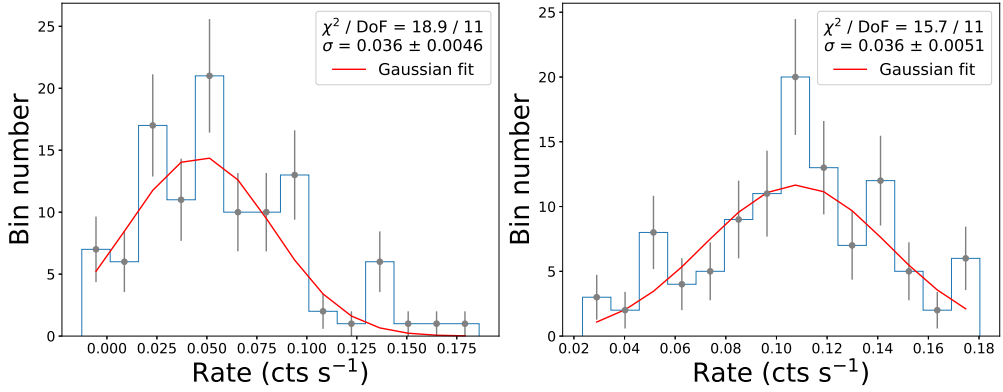} 
	\caption{Count-rate distributions for 4FGL~\tgt\ ({\it left}) and 
	the comparison source ({\it right}), constructed from the 100-s binned
	light curves. The best-fit Gaussian functions to each distributions
	are shown as red curves.
	\label{fig:blc}}
   \end{figure}
We constructed the count-rate distribution from the 100-s binned light curve 
of \tgt, which is shown in Figure~\ref{fig:blc}. To study any possible 
differences between it and that of the comparison source, which would help 
identify any possible variability of the target, the same distribution of
the comparison source was also constructed.
It can be seen that the comparison source was brighter, with the
distribution peak value $\mu$ at $\sim 0.11$\,cts\,s$^{-1}$. The value 
for \tgt\ is at $\sim 0.05$\,cts\,s$^{-1}$. We used a Gaussian function,
$\sim \exp[-(x-\mu)^2/2\sigma_g)]$, to fit the two
distributions, and obtained $\sigma_g \simeq 0.036\pm0.005$ 
for both \tgt\ and the comparison source. The $\chi^2$ values
were 18.9 and 15.7 respectively (with 11 degrees of freedom). From the fitting,
it can be seen that the distributions can generally be described with a Gaussian
function. For \tgt, a couple of data points, for example the one at 
$\sim 0.14$\,cts\,s$^{-1}$, are notably away from the best-fit Gaussian 
function, which is reflected from the slightly larger $\chi^2$ value. 
However, the deviations
are not significant. We thus conclude that the two sources have 
similar count-rate distributions.
\begin{figure}
   \centering
   \includegraphics[width=0.8\textwidth, angle=0]{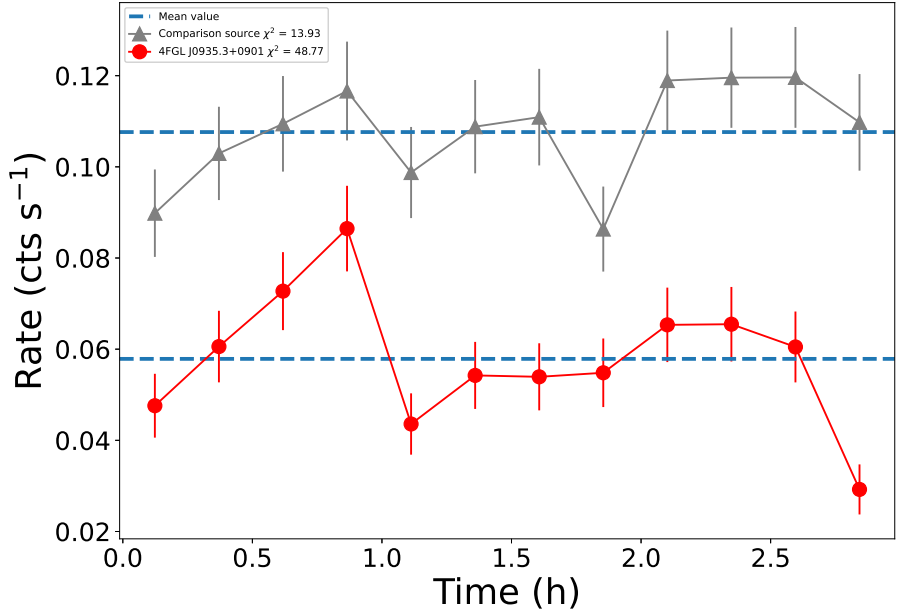} 
	\caption{890-s binned, background-subtracted light curves of 4FGL~\tgt\ (red) and 
	the comparison source (grey). 
	\label{fig:lc9}}
   \end{figure}

Because \tgt\ is a 2.5-hr binary, determined from optical photometry, we also
checked its relatively long time-duration variations. A 890-s binned light
curve, containing 12 data points, was constructed (Figure~\ref{fig:lc9}). 
We fit the light curve with a 
constant, and the resulting $\chi^2$ value was 48.77 (11 degrees of freedom). 
The same analysis was conducted for the comparison source, and the $\chi^2$ 
value was 13.93. The X-ray emission of 
\tgt\ was seemingly variable, by comparing with that of the comparison
source. However the light curves of the two sources appear to have a similar
temporal pattern. Their correlation coefficient was found to be $\sim0.3$, 
suggesting weak correlation. We checked our data analysis for possible 
background contamination, but no such effects were found. Therefore we conclude
that possible orbital variations were seen in the target.

\subsection{X-ray Spectrum}
\begin{figure}
   \centering
   \includegraphics[width=0.6\textwidth, angle=-90]{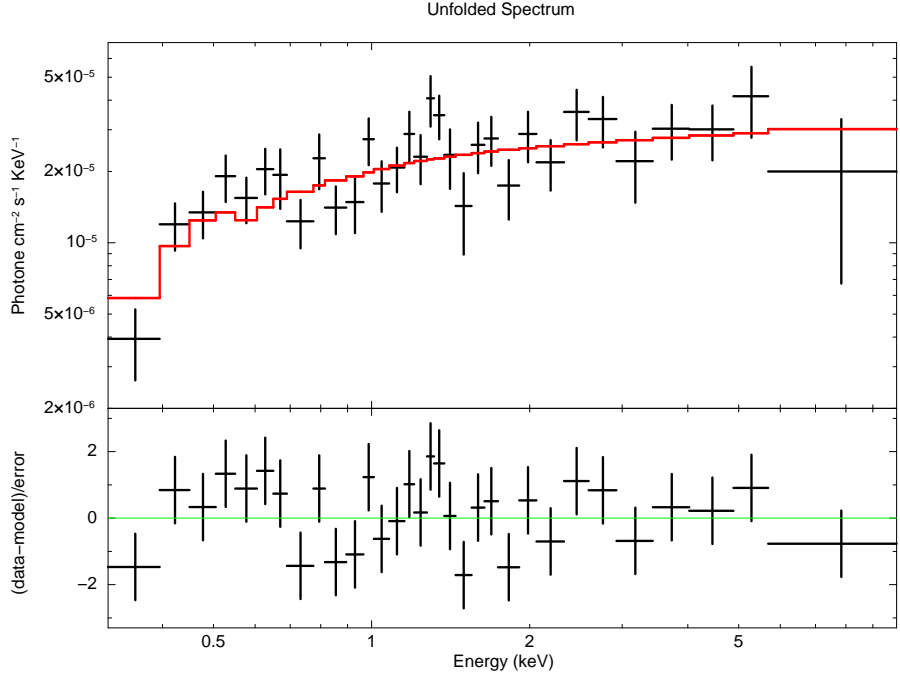} 
   \caption{X-ray spectrum of 4FGL~\tgt\ in 0.3--10\,keV. The spectrum can
	be well fit with an absorbed power law (red curve).}
   \label{fig:spec}
   \end{figure}

The 0.3--10\,keV spectrum of \tgt, combined from MOS and pn data,
is shown in Figure~\ref{fig:spec}. The spectrum can be well fit with an
absorbed power law. Using XSPEC (version 12.10.1f) for the fitting, we obtained 
a photon index of 1.88$^{+0.25}_{-0.22}$ and hydrogen column density 
$7.2^{+5.2}_{-4.5}\times 10^{20}$ cm$^{-2}$, where the errors are at a 90\%
confidence level (the reduced $\chi^2$ is 1.15 for 29 degrees of freedom).
The resulting 0.3--10\,keV absorbed flux is 
$ 1.23^{+0.09}_{-0.11} \times 10^{-13}$\,erg~s$^{-1}$\,cm$^{-2}$, 
and the unabsorbed is $ 1.45\pm0.15 \times 10^{-13}$\,erg~s$^{-1}$\,cm$^{-2}$.
We note that the Galactic hydrogen column density value \citep{hcd16}
is slightly lower than that we obtained, but within the uncertainty.

\subsection{FAST upper limit}

Because no pulsation signals were detected in our FAST observation, we derived
a flux density upper limit from the observation. The modified radiometer 
equation \citep{dtw+85,pjk+17} 
\begin{equation}
S_{min}=\frac{(S/N)\beta (T_{sys}+T_{sky})}{G\sqrt{n_{p}t_{obs}\Delta \nu}}\sqrt{\frac{W}{P-W}}\ \ \ ,
\end{equation}
was used, where the correction factor $\beta$ was assumed to be 1, 
the number of polarizations $n_{p}$ was 2, the observation time 
$t_{obs}$ was 20-min, and the bandwidth $\Delta\nu\simeq 400$\,MHz. 
Considering the fundamental performance of FAST described in \citet{jnl+20}, 
we adopted 20\,K for $T_{sys}+T_{sky}$ and  16\,K\,Jy$^{-1}$ for $G$, where 
$T_{sys}$ is the system noise temperature, $T_{sky}$ the sky temperature, and 
$G$ the gain of the telescope. Assuming a pulse duty cycle of 10\% ( 
$W$ and $P$ are the effective pulse width and the pulse period respectively), 
the upper limit value obtained was 0.4\,$\mu$Jy when S/N$=1$. 
Thus the upper limit on the flux density of 4FGL J0935.3$+$0901 was 
$\sim$4\,$\mu$Jy (when requiring S/N=10).

\section{Discussion}
\label{sec:disc}

We have observed 4FGL~\tgt\ with \xmm\ 
for the purpose of further identifying its nature, whether it could be a
candidate tMSP in the sub-luminous disk state.
From X-ray variability analysis, we did not find evidence for a
bimodal count-rate distribution, which would be expected from a tMSP
in the disk state. The \xmm\ observation confirmed the faintness of
the source at X-rays, as the obtained unabsorbed flux was 
$1.45\times 10^{-13}$\,erg~s$^{-1}$\,cm$^{-2}$, similar to the face value
of the X-ray flux obtained from short {\it Swift} observations reported
in \citet{wan+20}. The X-ray--to--$\gamma$-ray flux ratio now is updated to be
0.030$\pm0.007$. Both the variability analysis results and the flux ratio value 
indicate that the source is not a tMSP in the disk state; instead the both,
plus the X-ray photon index as well (the indices of redbacks are generally 
in a range of 1--2), are consistent with the properties
of redbacks in the pulsar state \citep{mil+20}. 

We found that the source's X-ray emission was possibly variable, 
while it is hard
to determine whether the variability was related to the 2.5-hr orbit because
of the limited observation time ($\simeq 3$ hrs).
Many redbacks in the pulsar state were found to be X-ray variable. They can show
clear orbital modulation (e.g., \citealt{bog+11,lin18}), flaring-like
variations (e.g., \citealt{hbt17,chb18}), or weak variations possibly related
to binary orbits (e.g., \citealt{lkh+16,lhs+18}). The X-ray variations likely
reflect intrabinary physical processes related to the pulsar winds, such
as the intrabinary shock \citep{bog+11} or magnetic reconnection in a striped
pulsar wind \citep{aln+18}. In this respect, the possible X-ray 
variation of
\tgt\ fits in the properties of redbacks. According to X-ray studies
of redbacks, their X-ray luminosities in the pulsar state are lower than
$10^{32}$\,erg\,s$^{-1}$ \citep{l14} and can be as low as 
$\sim 10^{31}$\,erg\,s$^{-1}$ \citep{str+19}. This would put \tgt\ at a 
distance of 
0.76\,kpc $(L_X/10^{31}\,{\rm erg\,s}^{-1})^{1/2}$, or $\leq 2.4$\,kpc when
requiring its X-ray luminosity $\leq 10^{32}$\,erg\,s$^{-1}$.

Considering 4FGL~\tgt\ as a redback in the disk-free, pulsar state, 
an explanation is needed for the double-peaked emission lines present in its 
optical emission, as such features are typically seen in spectra of accretion 
disks around compact stars.
The other possibility of arising from an outflow from the companion
and/or the intrabinary shock region
(see \citealt{swi+18} and references therein) should be the case. The emission
features of \tgt\ would be highly variable, which can be verified by 
spectroscopic observations. The existence of the intrabinary material 
could block the radio pulsed emission,
i.e., the often-seen eclipsing phenomenon in compact MSP binaries, 
and thus we did not detect the signals at the deep sensitivity
level of $\sim$4\,$\mu$Jy with FAST. Hopefully in the near future, further 
effort 
through multi-wavelength observations, including a radio observation 
covering the whole binary orbit (in order to avoid possible eclipsing phases),
may finally be able to provide a clear picture for this source.

\begin{acknowledgements}
This research is supported by the National Natural Science Foundation of 
China (11633007) and the Original
Innovation Program of the Chinese Academy of Sciences (E085021002).
J.V. thanks Prof. C.D. Ravikumar for the support and timely help.
\end{acknowledgements}


\end{document}